\documentclass[aps,preprint,graphicx,pra]{revtex4}
\usepackage[sort&compress]{natbib}
\usepackage{dcolumn,bm,graphicx,amsmath}

\begin{document}
\title{Simplified contact pseudopotential for anisotropic interactions of polarized particles
under harmonic confinement}

\author{  Andrei Derevianko}
\email{andrei@unr.edu}
\homepage{http://wolfweb.unr.edu/homepage/andrei/tap.html}

\affiliation{Physics Department, University of  Nevada, Reno, Nevada
89557}

\date{\today}

\begin{abstract}
Motivated by the recent progress in cooling and trapping polar molecules, we present
a simplified version of the rigorous contact pseudopotential for anisotropically-interacting polarized
particles [A. Derevianko, Phys. Rev. A 67, 033607 (2003); Phys. Rev. A 72, 039901(E) (2005)].
The simplifications are carried out for a practically important case of harmonically confined particles
described by sufficiently smooth wavefunctions.
The resulting contact pseudo-potential depends on the K-matrix of the underlying scattering process and
is represented as a sum over pairs of partial waves coupled by the collision.
The contribution of each pair of the coupled waves ($\ell$ and $\ell'$ ) involves a tensor
product of derivatives of orders $\ell$ and $\ell'$. The asymmetry in the derivatives reflects the
anisotropy of the original interaction potential: there is a preferential appearance of the
derivatives along the polarizing field.
\end{abstract}


%03.75.Hh Static properties of condensates;
%thermodynamical, statistical, and structural properties
% 34.10.+x General theories and models of atomic and
% molecular collisions and interactions (including statistical theories,
%transition state, stochastic and trajectory models, etc.)

\maketitle

\section{Introduction}
The intricacies of the  many-body problem are
rooted in inter-particle interactions  that lead to non-separable Hamiltonians.
Introducing pseudo-potentials, i.e., effective interactions that
are simpler than the original interactions, makes the problem more
tractable.  In particular, due to their
compelling efficiency, the {\em contact} (zero-range) pseudo-potentials~\cite{DemOst88}
have enjoyed a
remarkable success in a number of
applications, the physics of ultracold atomic gases being one of the most recent
examples.  In particular, properties of
Bose-Einstein condensates
may be well understood just in terms of a contact potential (Fermi~\cite{Fer36}), with
its strength determined by a single scattering parameter -- scattering length \cite{DalGioPit99,Leg01}.

Due to their importance in ultracold atomic physics,
there has been a recent resurgence of interest in the pseudo-potential approximation, see, e.g.,
Refs.~\cite{RotFel01,BolTieJul02,BluGre02,Der05HY,IdzCal06,Pri06,ReiSilSto06,ColMasPet07}.
All of the enumerated works
focused on particles that predominantly interact via
spherically-symmetric forces. Here, motivated by recent experimental
developments, we focus on anisotropically interacting atoms and molecules.
Indeed, much of the present experimental efforts on cooling and trapping
of atomic particles  turned to ``unconventional'' (non-alkali) species:
molecules~\cite{specIssuePolMol04}  and atoms with complicated open-shell structure~\cite{KimFriKat97,WeideCKim98,BelStuLoc99}.
Most of these species possess internal multipolar moments, such as electric and magnetic dipoles, quadrupole moments, etc. As a result, polarized
atoms and molecules interact via anisotropic  forces: the interactions depend on orientation
of the inter-particle axis with respect to the orientation of the internal moments.

Previously,
I derived a rigorous contact pseudopotential for anisotropically interacting species~\cite{Der03}.
While being general, the rigorous prescription is hardly necessary for smoothly behaving
wavefunctions (mathematically, we require the existence of the Fourier transform
 (i.e., the momentum representation) of the wavefunction and  all of its derivatives).
 Moreover,
most of the cold-atom experiments are carried out in a presence of
harmonic trapping potentials. These two observations facilitate
simplifications of  the rigorous pseudopotential.
In particular, we show that integro-differential operators entering the rigorous pseudopotential
may be replaced by a mathematically simpler combination of derivatives.

One may distinguish between two classes of applications of  pseudopotentials:
(i) determining exact few-body solutions and (ii) usage of pseudopotentials within a many-body mean-field  framework. In the former case one needs to carefully deal with regularization issues and the rigorous pseudo-potential
is required. In the later, many-body, case,
the functional space is restricted and the regularization is not needed. The results of this paper are geared towards the many-body applications of the pseudopotential method.

This paper is organized as follows. First, In Section~\ref{Sec:Setup}, we formalize
the problem and recapitulate results for the rigorous pseudopotential of Ref.~\cite{Der03}.
In Section~\ref{Sec:HOargument}, we consider simplifying conditions of the collision
process in a harmonic trap. Before proceeding to
general derivation in Section~\ref{Sec:GeneralCase}, in Section~\ref{Sec:truncatedDipolar}
we illustrate the involved techniques by simplifying a truncated dipolar pseudopotential and  the
contact Fermi interaction. In Sections~\ref{Sec:Isotropic} and \ref{Sec:Born},
we consider two limiting cases of the isotropic interactions and anisotropic
interactions in the Born approximation. Finally, the summary is given in Section~\ref{Sec:Summary}.

\section{Problem setup}
\label{Sec:Setup}
While the formal setup of the problem is identical to that of Ref.~\cite{Der03},
below I would like to recapitulate several points as they relate to more recent
developments in ultracold physics with polarized heteronuclear molecules.

In a typical cold-molecule setup,
a dilute gas of polar molecules forms a cloud in an external harmonic trapping
potential. Orientation of molecular dipoles $D$ is fixed by application of a polarizing electric field
(otherwise, molecular rotations would average dipole moments to zero).
At large inter-molecular separations,
  the molecular interactions have the dipolar character,
\begin{equation}
 V(\mathbf{r}) \rightarrow
 \frac{D^2}{r^3} (3 \cos^2 \theta -1)\, . \label{Eq:Vdd}
\end{equation}
Here $\theta$ is the angle between the inter-molecular (collision) axis $\mathbf{r}$ and the
polarizing field (it is assumed to be directed along the z-axis). Clearly the interaction is anisotropic: if the collision axis
is aligned with the external field, the dipoles attract each other, and if
$\mathbf{r}$ is perpendicular to the field, the interaction is repulsive.

Dipolar interactions are also important for the BEC of highly-magnetic chromium
atoms~\cite{GriWerHen05}.  Again the species are polarized by applying an external (magnetic)
field.  The atoms are trapped in the lowest-energy Zeeman sublevel; the depolarizing transitions to the
upper-energy levels are forbidden energetically.

While most of our discussion focuses on the dipolar interactions, the final result
will be also valid for a general case of  axially-symmetric interactions.
We will parameterized such interactions as
\begin{equation}
   V(\mathbf{r}) = V_{L}(r) P_{L}(\cos \theta) \,, L \, \text{is even} \, ,
   \label{Eq:VL-general}
\end{equation}
where $P_{L}$ is the Legendre polynomial. Due to the axial symmetry, $L$ is an even integer. In particular,
for isotropic interactions, $L=0$.
For dipolar interaction (\ref{Eq:Vdd}), $L=2$, $V_2(r) = -2D^2/r^3 $.

Constructing the pseudopotential in the Huang-Yang approach~\cite{HuaYan57}, essentially boils down to finding an effective
contact interaction which, at sufficiently large separations,
reproduces the exact solutions of the Schr\"{o}dinger  equation with the full original potential.
To this end one requires knowing the exact solutions of the scattering problem with the full potential
in the asymptotic region. The relevant radius of the asymptotic region may be estimated by comparing centrifugal term in the corresponding radial equations with
the multipolar contribution. Parameterizing $V_L(r) = C_k/r^k$, the radius is
\begin{equation}
  \bar{R}_{k} = \left( \frac{M}{\hbar^2} C_{k} \right)^{1/(k-2)} \,.
 \label{Eq:Rmulti}
\end{equation}
While the pseudopotential would reproduce the solutions for $r \gg \bar{R}_{k}$,
it is worth emphasizing that the scattering solution depends on the  entire interaction potential:
as molecules approach each other, the electronic clouds start to overlap, and the
interactions substantially depart from the dipolar form (\ref{Eq:Vdd}). In addition,
the anisotropic interactions  couple various
partial waves: for example, the dipolar interaction nominally couples partial wave $\ell$ with partial
waves $\ell \pm 2$.  In other words, one has to solve a multi-channel scattering problem.

The exact linear combinations of the regular and irregular free-particle solutions in the asymptotic region are naturally expressed in terms of the K-matrix (reactance
matrix), $\mathcal{K}_{\ell m}^{ \ell' m'}$, which characterizes couplings
between $(\ell m)$ and $(\ell' m')$ partial waves. To formalize the concept of the K-matrix, we consider
scattering wavefunction $\Phi\left(  \mathbf{r}\right)$ for a pair of particles colliding
with the relative linear momentum of
$\hbar k$.
In the asymptotic region it may
be expanded using free-particle solutions
\begin{equation}
\Phi\left(  \mathbf{r}%
\right) \rightarrow \sum_{lm}\left(  \alpha_{lm}\, j_{l}\left(  kr\right)  -\beta
_{lm}n_{l}\left(  kr\right)  \right)  Y_{lm}\left( \hat{ \mathbf{r}} \right)
\, ,\label{Eq:asmpt}%
\end{equation}
where $j_{l}\left(  kr\right)  $ and $n_{l}\left(  kr\right)  $ are the spherical
Bessel and Neumann functions respectively and $\alpha_{lm}$ and $\beta_{lm}$
are integration constants. These constants are related
requiring the solution $\Phi$ of the Schrodinger equation with  full potential to be regular at $r=0$,
\begin{equation}
\beta_{lm}=\sum_{l^{\prime}m^{\prime}}\mathcal{K}_{lm}^{l^{\prime}m^{\prime}%
}\alpha_{l^{\prime}m^{\prime}} \label{Eq:betaKalpha} \, .
\end{equation}
This relation defines the K-matrix.

The anisotropic pseudo-potential~\cite{Der03}   reads
\begin{equation}
\hat{V}_{ps}\Psi(\mathbf{r})=-\frac{\hbar^{2}}{M}\sum_{ll^{\prime};mm^{\prime}}\frac
{\mathcal{K}_{lm}^{l^{\prime}m^{\prime}}}{k^{l+l^{\prime}+1}}\frac{\left(  2l+1\right)
!}{2^{l}l!}Y_{lm}\left(  \mathbf{\hat{r}}\right)  \frac{\delta\left(
r\right)  }{r^{l+2}} \, \left( \Pi_{l^{\prime}m^{\prime}}\Psi \right) \, ,
\label{Eq:regPseudo}
\end{equation}
where $\delta\left(
r\right)$ is the Dirac $\delta$-function and  the  projection operator is defined as
\begin{equation}
\Pi_{l^{\prime}m^{\prime}}\Psi\equiv\frac{1}{2^{l^{\prime}}l^{\prime}!}%
\lim_{r\rightarrow0}\left(  \frac{d}{dr}\right)  ^{2l^{\prime}+1}r^{l^{\prime
}+1}\int Y_{l^{\prime}m^{\prime}}^{\ast}\left(  \Omega\right)  \Psi\left(
\mathbf{r}\right)  d\Omega \, . \label{Eq:Projector}
\end{equation}

The matrix element
of the  pseudopotential~(\ref{Eq:regPseudo}) between two plane waves $\langle
\mathbf{r}|\mathbf{k}\rangle=\left(  2\pi\right)  ^{-3/2}e^{i\mathbf{k}%
\cdot\mathbf{r}}$ is given by
\begin{equation}
\bar{v}\left(  \mathbf{k,k}^{\prime}\right)  =-\frac{\hbar^{2}}{M} \frac{2}{\pi} \sum_{ll^{\prime};mm^{\prime}}i^{l^{\prime}-l}\frac{K_{lm}^{l^{\prime
}m^{\prime}}\left(  k^{\prime}\right)  }{k^{\prime}}\left(  \frac{k}%
{k^{\prime}}\right)  ^{l}Y_{l^{\prime}m^{\prime}}^{\ast}\left(  \mathbf{\hat
{k}}^{\prime}\right)  Y_{lm}\left(  \mathbf{\hat{k}}\right) \,.
\label{Eq:kernel}
\end{equation}

In the following we assume that the global many-body properties of the quantum gas
can be described by well-behaved wavefunctions. We will base our derivation on
the momentum-space kernel Eq.~(\ref{Eq:kernel}). This leads to the requirement that
the momentum representation of the wavefunction and of all of its
derivatives must exist. This constraint follows from the fact that one needs
to interchange differentiation and integration operators in
Eq.(\ref{Eq:Projector}) while arriving to Eq.(\ref{Eq:kernel}) for a momentum-space wavefunction.
This constraint, for example, discards wavefunctions diverging as $1/r$ at the origin.

Let me explicitly address a somewhat confusing point of dipolar interactions being of ``long-range''
character. Can the contact pseudopotential approximate an action of the
``long-range'' dipolar interaction?
The use of the contact pseudopotential is mathematically justified
 because there  is a radius $\bar{R}_3$,  Eq.(~\ref{Eq:Rmulti}), where the centrifugal
 forces overtake the dipolar interaction. Beyond $\bar{R}_3$ the wavefunction
 no longer ``feels'' the dipolar interaction, as reflected by Eq.(~\ref{Eq:asmpt}).
 In this sense, the dipolar interactions are not ``long-range'' but
 rather ``long(ER)-range''. The true anisotropic ``long-range'' potentials would
 have $1/R^k$ tail,  where $k \le 2$. Notice that in this discussion, I have
 used the fact that the anisotropic potentials do not directly contribute to
 diagonal part of the $\ell=0$ radial equation void of the centrifugal contribution.

\section{Simplifications for harmonically-confined particles}
\label{Sec:HOargument}
Now, under simplifying assumption of harmonic trapping, I transform the
momentum-space expression~(\ref{Eq:kernel}) back into the coordinate space.  We write for a
matrix element of the pseudopotential (cf. Ref.~\cite{Omo77} for Rydberg atoms)
\begin{eqnarray}
\langle\psi|\hat{V}_\mathrm{ps}|\psi\rangle&=&{\left(  2\pi\right)^{-3}}
\int
d\mathbf{k}d\mathbf{k}^{\prime}d\mathbf{r}d\mathbf{r}^{\prime} \times
\nonumber \\
&&
\psi^{\ast}\left(  \mathbf{r}\right)
e^{-i\mathbf{k}\cdot\mathbf{r}}
\bar{v}\left( \mathbf{k}, \mathbf{k}^{\prime}\right)
 \psi^{\ast}\left(  \mathbf{r}^{\prime}\right)  e^{i\mathbf{k}^{\prime}\cdot
\mathbf{r}^{\prime}}
 \,
 .
\label{Eq:Meltransform}%
\end{eqnarray}
Below we simplify this expression by noticing that
only certain values of $|\mathbf{k}|$ and $|\mathbf{k}^{\prime}|$
contribute to this integral.

Experimentally,
the collisions occur in
the presence of a harmonic trapping potential
\begin{equation}
U\left(  \mathbf{r}\right)
=\frac{1}{2}M\left(  \omega_{x}^{2}x^{2}+\omega_{y}^{2}y^{2}+\omega_{z}
^{2}z^{2}\right) \, ,
\label{Eq:Uharmonic}
\end{equation}
where $\omega_i$ are the trapping frequencies.
For two harmonically-confined particles the center-of-mass
and relative motions decouple and the Hamiltonian for the relative motion
reads
\[
H_{r}= \frac{p_{12}^{2}}{  2\mu }  +\left(  \frac{\mu}{M}\right)
U\left(  \left\vert \mathbf{r}_{12}\right\vert \right)  +V\left(
\mathbf{r}_{12}\right) \, ,
\]
where $V\left(  \mathbf{r}_{12}\right)  $ is the
full interaction potential between the particles and $\mu=M/2$ is the reduces
mass. In the
stationary problem, we solve the eigenvalue equation $H_{r}\psi\left(
\mathbf{r}_{12}\right)  =E_{r}\psi\left(  \mathbf{r}_{12}\right)  $, $E_{r}$
being the energy of the relative motion. Ref.~\cite{BluGre02,BolTieJul02}
presented a numerical
comparison of solutions of this equation with the full molecular
potential and its pseudopotential representation (for isotropic $s$-wave
scattering).
Their main conclusion is that the pseudopotential description
holds as long as the  range of the potential is much smaller than the
oscillator length $L_{i}=(\hbar/M\omega_{i})^{1/2}$.
Ref.~\cite{BluGre02}  also observed that
for
velocity-dependent potentials the relevant collision momentum in $V_\mathrm{ps}$ is
$\hbar^{2}k_{c}^{2}/(2\mu)=E_{r}$. Indeed,
the collision process occurs at
$\left\vert \mathbf{r}_{12}\right\vert $ much smaller than the harmonic length.
In this region
$U\left(  \left\vert \mathbf{r}_{12}\right\vert \right)  \approx0$ and
the kinetic energy is  $E_{r}$.

Recently, there was a study~\cite{KanBohBlu07} of the validity of the {\em anisotropic}
pseudopotential approach
for dipolar interactions~\cite{Der03} in a harmonic trap. Similarly to the isotropic case\cite{BluGre02,BolTieJul02},
these authors found that the pseudopotential description remains accurate as long as
the dipolar length $\bar{R}_3$ is smaller than the characteristic length of the trapping potential.
Their numerical experiments also support the discussed choice of the collision momentum $k_c$.

Based on the above comparison with the exactly-solvable two-body case, we require that the multipolar lengths, Eq.~(\ref{Eq:Rmulti}), are much smaller than the
harmonic lengths, $\bar{R}_{L+1} \ll \min L_i $, i.e., the pseudo-potential treatment
is valid in the region where the wavefunction remains essentially flat.
We notice in passing, that the more rigorous consideration should include criteria related
to the scattering parameters of the problem, such as the scattering length~\cite{KanBohBlu07}.
Under this condition,
the interparticle interactions have a negligible effect on the
energies of the harmonic motion, as the characteristic multipolar interaction energy
is much smaller than the excitation energy of the harmonic potential and
we identify
$E_r \approx \hbar (n_x \omega_x + n_y \omega_y +n_z \omega_z + 3/2)$, $n_i =0,1,2, \ldots$.
For the ground state of the harmonic oscillator
the above discussion leads to the value of the collision momentum of
\begin{equation}
k_{c}^{2} \approx 3/2~M \bar{\omega}/\hbar \, , \label{Eq:kcChoiceGroundState}
\end{equation}
  with $\bar{\omega}=\sum_i
\omega_i/3$ being the average of the three trapping frequencies.
The generalization to the excited states is straightforward.

We also notice that for a many-particle case, due to mean-field effects,
the effective potential felt by the particles may deviate from
the external harmonic trapping potential, Eq.(\ref{Eq:Uharmonic}).
In this case the mean-field wavefunction needs to be fitted to
a corresponding harmonic oscillator wavefunction and the $k_c$ has
to be consistently redefined based on the fitting. Indeed,
most of the variational  studies of  BECs employ the Gaussian
ansatz for the trial wavefunction, see for example, Ref.~\cite{PetSmi02}.
In this sense, the choice of $k_c$
would be self-consistent in such variational studies. Otherwise,
the suggested fitting is an approximate prescription.

Returning to the evaluation of the integral~(\ref{Eq:Meltransform}),
we see that the relevant
contributions are accumulated at $\left\vert \mathbf{k}\right\vert =\left\vert
\mathbf{k}^{\prime}\right\vert =k_{c}$. In other words,  we consider
only the ``on-shell'' collisions, with the radius of the shell determined
by the energy of the harmonic motion. This situation is similar to the interaction
of Rydberg electrons (of well-defined kinetic energy) with atomic particles,
considered by \citet{Omo77}.
The present derivation may be considered as a generalization of his approach to
anisotropically interacting particles. We will recover the Omont's result
as a limiting case of isotropic interactions in Section~\ref{Sec:Isotropic}.

\section{Truncated dipolar pseudopotential}
\label{Sec:truncatedDipolar}
Before presenting the general analysis, I will demonstrate
the procedure and discuss some subtleties for a simpler case of the
truncated dipolar pseudopotential~\cite{Der03}.
In this model, the entire summation over the coupled partial waves in the momentum-space kernel~(\ref{Eq:kernel})
is limited to the $s-s$ and $s-d$
couplings,
\[
\bar{v}\left(  \mathbf{k},\mathbf{k}^{\prime}\right)   \approx \frac{1}{2\pi^{2}%
}\frac{\hbar^{2}}{M}\left(  a_{ss}-a_{sd}~\mathcal{F}\left(  \mathbf{k},\mathbf{k}%
^{\prime}\right)  \right) \, ,
\] with
\[
\mathcal{F}\left(  \mathbf{k},\mathbf{k}^{\prime}\right)  = \sqrt{5}\left\{
P_{2}\left(  \cos\theta_{k}\right)  +\left(  k/k^{\prime} \right)
^{2}P_{2}\left(  \cos\theta_{k^{\prime}}\right)  \right\}  \, ,
\]
where $\theta_{k}$ and $\theta_{k^{\prime}}$ are angles between the
polarizing field  and $\mathbf{k}$ and
$\mathbf{k}^{\prime}$. Parameter $a_{ss}$ is the conventional scattering length
\[
 a_{ss} = - \mathcal{K}^{00}_{00}(k) /k \, ,
\]
and $a_{sd}$ is the so-called ``off-diagonal'' scattering length,
characterizing the strength of anisotropic coupling between the s and d waves
\[
 a_{sd} = - \mathcal{K}^{20}_{00}(k) /k \, .
\]
As  $k\rightarrow 0$, both expressions approach finite values.

Let us consider the isotropic ($a_{ss}$) term first. We expect to recover
the original Fermi contact interaction, $4\pi\, \hbar^{2}/ M \, a_{ss}\, \delta\left(
\mathbf{r}\right)$, as the result of the analysis.
\begin{align*}
\frac{1}{\left(  2\pi\right)  ^{3}}\int d\mathbf{k}d\mathbf{k}^{\prime
}e^{i\mathbf{k}^{\prime}\cdot\mathbf{r}^{\prime}}\bar{v}\left(  \mathbf{k}%
^{\prime},\mathbf{k}\right)  e^{-i\mathbf{k}\cdot\mathbf{r}}  &  =\frac
{1}{\left(  2\pi\right)  ^{3}}\frac{1}{2\pi^{2}}\frac{\hbar^{2}}{M}a_{ss}\int
d\mathbf{k}\int d\mathbf{k}^{\prime}e^{i\mathbf{k}^{\prime}\cdot
\mathbf{r}^{\prime}}~1e^{-i\mathbf{k}\cdot\mathbf{r}}~=\\
&  \frac{1}{2\pi^{2}}\frac{\hbar^{2}}{M}a_{ss}\left(  2\pi\right)  ^{3}%
\delta\left(  \mathbf{r}\right)  \delta\left(  \mathbf{r}^{\prime}\right) \, ,
\end{align*}
where we used the identity $
\int d^{3}k~\exp[i\mathbf{k}\cdot\mathbf{r}]=\left(  2\pi\right)  ^{3}%
\delta\left(  \mathbf{r}\right)$. Then
\[
\langle\psi|V_{ps}|\psi\rangle=\frac{1}{2\pi^{2}}\frac{\hbar^{2}}{M}%
a_{ss}\left(  2\pi\right)  ^{3}\int\int\delta\left(  \mathbf{r}\right)
\delta\left(  \mathbf{r}^{\prime}\right)  \psi^{\ast}\left(  \mathbf{r}%
^{\prime}\right)  \psi\left(  \mathbf{r}\right)  drdr^{\prime}=\frac{1}%
{2\pi^{2}}\frac{\hbar^{2}}{M}a_{ss}\left(  2\pi\right)  ^{3}\psi^{\ast}\left(0\right)  \psi\left(  0\right)
\]
From here we read
\begin{equation}
V_{ps}\left(  \mathbf{r}\right)  =4\pi\frac{\hbar^{2}}{M}a_{ss}\delta\left(
\mathbf{r}\right) \, ,
\end{equation}
which is, indeed, the Fermi contact interaction.

Now we proceed to simplifying the dipolar ($a_{sd}$) term. We focus on the following integral
\[
I=\int d\mathbf{k}\int d\mathbf{k}^{\prime}e^{i\mathbf{k}^{\prime}%
\cdot\mathbf{r}^{\prime}}~P_{2}\left(  \cos\theta_{k}\right)  e^{-i\mathbf{k}%
\cdot\mathbf{r}}~ \, .
\]
The resulting operator form of the pseudopotential depends on the choice of the representation of the Legendre polynomial
$P_{2}\left(  \cos\theta_{k}\right)  =-\frac{1}{2}+\frac{3}{2}\cos^{2}%
\theta_{k}$. We may express the polynomial as
\[
P_{2}\left(  \cos\theta_{k}\right)  =-\frac{1}{2}+\frac{3}{2}\frac{k_{z}^{2}%
}{k^{2}} \, ,
\]
or, equivalently, as
\[
P_{2}\left(  \cos\theta_{k}\right)  =-\frac{1}{2}\frac{1}{k^{2}}\left(
k_{x}^{2}+k_{y}^{2}+k_{z}^{2}\right)  +\frac{3}{2}\frac{k_{z}^{2}}{k^{2}} \, .
\]
While computing the integral, we evaluate the $|\mathbf{k}|$-dependent functions
at the ``on-shell'' value of $|\mathbf{k}|=k_c$ and pull them out of the integration.
We also use $
\int k_{z}^{2} \exp(i \mathbf{k} \cdot \mathbf{r} ) \, d\mathbf{k}    =\left(  \frac{1}{i}\right)  ^{2}\frac{\partial^{2}}{\partial
z^{2}}\int \exp(i \mathbf{k} \cdot \mathbf{r} ) \, d\mathbf{k}
= \left(  2\pi\right)  ^{3}\frac
{1}{i^{2}}\frac{\partial^{2}}{\partial z^{2}}\delta\left(
\mathbf{r}\right)$.
With the first combination we obtain
\[
I=-\frac{1}{2}\delta\left(  \mathbf{r}^{\prime}\right)  \delta\left(
\mathbf{r}\right)  -\frac{3}{2}\frac{1}{k_{c}^{2}}\delta\left(  \mathbf{r}%
^{\prime}\right)  \frac{\partial^{2}}{\partial z^{2}}\delta\left(
\mathbf{r}\right) \,
\]
and with the second combination%
\[
I=\delta\left(  \mathbf{r}^{\prime}\right)  \frac{1}{k_{c}^{2}%
}\left(  \frac{1}{2}\Delta-\frac{3}{2}\frac{\partial^{2}}{\partial z^{2}%
}\right)  \delta\left(  \mathbf{r}\right) \, .
\]
Both combinations are equivalent for harmonic-oscillator wavefunctions.
The equivalence depends on the definition of the collision momentum $k_c$;
it has to be chosen consistently.

In the first case we  arrive at the pseudopotential
\begin{eqnarray}
\hat{V}_\mathrm{ps}\left(  \mathbf{r}_{12}\right) & \approx & 4\pi\frac{\hbar^{2}}{M}%
\delta\left(  \mathbf{r}_{12}\right) \times  \label{Eq:PSgaugeZero} \\
&& \left\{  a_{ss}+\sqrt{5}a_{sd}\left[
1+\frac{3}{2}\frac{1}{k_{c}^{2}}\left(  \overleftarrow{\frac{\partial^2}{\partial_{z_{12}}^{2}}
}+\overrightarrow{
\frac{\partial^2}{\partial_{z_{12}}^{2}}}\right)  \right]  \right\}  ,
\nonumber%
\end{eqnarray}
and in the second case to the equivalent pseudopotential
\begin{eqnarray}
\hat{V}_\mathrm{ps}\left(  \mathbf{r}_{12}\right) & \approx & 4\pi\frac{\hbar^{2}}{M}%
\delta\left(  \mathbf{r}_{12}\right) \times   \label{Eq:PSTensorGauge}  \\
&& \left\{  a_{ss}+\sqrt{5}a_{sd}
\frac{1}{k_{c}^{2}}
\left[
-\frac{1}{2} \left( \overleftarrow{\Delta} + \overrightarrow{\Delta} \right)
  +\frac{3}{2}\left(  \overleftarrow{\frac{\partial^2}{\partial_{z_{12}}^{2}}
}+\overrightarrow{
\frac{\partial^2}{\partial_{z_{12}}^{2}}}\right)  \right]  \right\}  .
\nonumber%
\end{eqnarray}
In these expressions, the operator $\overleftarrow{.}$
acts on the bra and $\overrightarrow{.}$ operates on the
ket.
In both cases  the
dipolar contribution to $\hat{V}_\mathrm{ps}$ breaks into two parts:
isotropic and anisotropic terms. The anisotropy is imposed by the
polarizing field (the derivatives in the anisotropic contributions are taken along the field).
The long-range character of the dipolar interactions is manifested through
$k_c^2$ which characterizes the entire trapping potential.

Being equivalent, which representation, Eq.(\ref{Eq:PSgaugeZero}) or Eq.(\ref{Eq:PSTensorGauge}), is more accurate in approximate calculations? A similar
question appears in choosing the gauge-dependent operators (e.g., velocity-gauge and length-gauge of the
dipole operator) when carrying out computations involving radiation fields in atomic physics or quantum electro-dynamics.
When carrying out approximate calculations the results would differ.
The advantage of representation~(\ref{Eq:PSgaugeZero})
is that the isotropic part of the dipolar operator participates on an equal footing with the Fermi contribution: the isotropic dipolar
contribution merely renormalizes the traditional spherically-symmetric
pseudopotential; this may be beneficial in interpreting the results of the calculations.
The advantage  of the second representation~(\ref{Eq:PSTensorGauge})
is that it is the zeroth component of the rank 2 irreducible spherical tensor (we will discuss
this in detail in the following Section).
As such its expectation value manifestly vanishes for {\em any} spherically-symmetric wavefunction,
while the expectation value of ~(\ref{Eq:PSgaugeZero}) vanishes only for the  spherically-symmetric states of the harmonic oscillator.

Since the properties of dipolar quantum gasses are usually studied as a function of the
aspect ratio of the trapping potential, see e.g., ~\cite{YiYou00,SanShlZol00}, the tensorial representation seems to have a computational advantage. Another
advantage is that the resulting expressions for higher partial waves are more concise. Due to these considerations, below I will treat the general case
using techniques of irreducible tensor operators.

\section{General case}
\label{Sec:GeneralCase}
In this Section we carry out simplifications of the rigorous pseudopotential~(\ref{Eq:regPseudo})
for an arbitrary potential and all partial waves. The derivation essentially
follows the steps for the truncated dipolar pseudopotential of the preceding
Section. I will use several formulae from the quantum theory of angular momentum. The notation
and phase convention follow that of the standard compilation~\cite{VarMosKhe88}.
In particular, normalized spherical harmonics (C-tensors) are defined as  $C^{l}_{m}(\hat{\mathbf{r}})=\sqrt{\frac{4\pi}{2l+1}} \;Y_{lm}(\hat{\mathbf{r}})$.
The tensor product of irreducible  tensor operators (such as C-tensors) is
expressed in terms of their spherical components as
\begin{equation}
\left\{  A^{\left(  j_{1}\right)  }\otimes~B^{\left(  j_{2}\right)  }\right\}
_{JM}=\sum_{m_{1}m_{2}}C_{j_{1}m_{1}j_{2}m_{2}}^{JM}A_{m_{1}}^{\left(
j_{1}\right)  }B_{m_{2}}^{\left(  j_{2}\right)  } \, ,
\end{equation}
where $C_{j_{1}m_{1}j_{2}m_{2}}^{JM}$ are the Clebsh-Gordan coefficients.
Beyond streamlined notation, there are several advantages to operating in
terms of the tensor products, e.g.,  angular selection rules and
a suitability for applying the Wigner-Eckart theorem.
We will also make use of the following representation of the C-tensor
\begin{equation}
C_{m}^{l}\left(  \theta,\varphi\right)  =\frac{1}{r^{l}}\left[  \left(
l-m\right)  !\left(  l+m\right)  !\right]  ^{1/2}\sum_{p,q,r}\frac{1}%
{p!q!r!}\left(  -\frac{x+iy}{2}\right)  ^{p}\left(  \frac{x-iy}{2}\right)
^{q}z^{r} \, , \label{Eq:C-tensor}
\end{equation}
where $p,q,$ and $r$ are non-negative integers satisfying $p+q+r=l$ and
$p-q=m$.

In addition we will use general properties of the K-matrix $\mathcal{K}_{lm}^{l'm'}$.
First of all, the polarizing field imposes the axial symmetry on the collision process.
This symmetry leads to the conservation of the projection of the angular momentum on the
direction of the polarizing field (z-axis) and to the selection rule $m=m'$.
Parity conservation requires that only the partial waves of the same parity are
coupled: for example, the $p$-waves are uncoupled from the $s$ and $d$ waves.
Formally,
\[
  \mathcal{K}_{lm}^{l'm'} = \delta_{m,m'} \frac{1}{2} \, (1+(-1)^{l+l'}) \, \mathcal{K}_{lm}^{l'm} \,.
\]
We also relate $\mathcal{K}_{l,-m}^{l',-m'} =  \mathcal{K}_{l,+m}^{l',+m'}$. Finally, the
time-reversal symmetry leads to $\mathcal{K}_{lm}^{l'm'} =  \mathcal{K}^{lm}_{l'm'}$.
Again it is worth reminding that we limit our consideration to axially-symmetric components
of the full potential~(\ref{Eq:VL-general}), where
$L$ is {\em even}, and the enumerated properties can be derived by a direct examination
of the radial scattering equations (see, e.g., appendix of Ref.~\cite{Der03}).

We may rewrite the momentum-space kernel, Eq.~(\ref{Eq:kernel}) as an expansion over tensor products
of normalized spherical harmonics%
\begin{equation}
\bar{v}_{k_{c}}\left(  \mathbf{k,k}^{\prime}\right)     =
\frac{1}{(2 \pi)^3} \, \sum_{J~even~}%
\sum_{ll^{\prime}}T_{J}\left(  ll^{\prime};k_{c}\right)  i^{l^{\prime}%
-l}\left\{  C^{l}\left(  \mathbf{\hat{k}}\right)  \otimes C^{l^{\prime}%
}\left(  \mathbf{\hat{k}}^{\prime}\right)  \right\}  _{J,M=0} \, ,
\end{equation}
with the coupling coefficient
\begin{equation}
T_{J}\left(  ll^{\prime};k_{c}\right)     =- 4 \pi \frac{\hbar^{2}}{M}
\left[  (2l+1)(2l^{\prime}+1)\right]  ^{1/2}\sum_{m}\left(
-1\right)  ^{m}C_{l,m;l^{\prime},-m}^{J,M=0}\frac{K_{lm}^{l^{\prime}m}\left(
k_{c}\right)  }{k_{c}} \, .
\label{Eq:Tcoupling}
\end{equation}
In the sum, the compound angular momentum $J$ runs over all even integers in the
range $\left\vert l-l^{\prime}\right\vert \leq J\leq l+l^{\prime}$.
Since both $l$ and $l^{\prime}$ are simultaneously even or odd (parity
selection rule), the sum is limited to  {\em even} values of $J$.
In addition, following the arguments of
Section~\ref{Sec:HOargument} we have evaluated the kernel at the on-shell value of $\left\vert
\mathbf{k}\right\vert =\left\vert \mathbf{k}^{\prime}\right\vert =k_{c}$.
 Notice that only the $M=0$ component of the compound tensor participates in
the expansion - this is due to the axial symmetry of the collision process.

Now we turn to evaluating the integral over momenta in Eq.(\ref{Eq:Meltransform}). We expand the
C-tensors according to Eq.(\ref{Eq:C-tensor}) and use the properties of the Dirac $\delta
$-function as in Section~\ref{Sec:truncatedDipolar}. On recombining the resulting expansion we arrive at%
\begin{equation}
\langle\psi|V_{ps}|\psi\rangle=\sum_{J~even~}\sum_{ll^{\prime}}T_{J}\left(
ll^{\prime};k_{c}\right)  \lim_{\mathbf{r}\rightarrow0}\psi^{\ast}\left(
\mathbf{r}\right)  \left\{  C^{l}\left(  \frac{\overleftarrow{\mathbf{\nabla}%
}}{k_{c}}\right)  \otimes C^{l^{\prime}}\left(  \frac{\overrightarrow
{\mathbf{\nabla}}}{k_{c}}\right)  \right\}  _{J,0}\psi\left(  \mathbf{r}%
\right) \, ,
\label{Eq:VpsGeneralMel}
\end{equation}
or in the operator form%
\begin{equation}
V_{ps}\left(  \mathbf{r}\right)  =\sum_{J~even~}\sum_{ll^{\prime}}T_{J}\left(
ll^{\prime};k_{c}\right)  \delta\left(  \mathbf{r}\right)
 \left\{  C^{l}\left(  \frac{\overleftarrow
{\mathbf{\nabla}}}{k_{c}}\right)  \otimes
C^{l^{\prime}}\left(  \frac{\overrightarrow{\mathbf{\nabla}}}{k_{c}}\right)
\right\}  _{J,0}  \, .
\label{Eq:VpsGeneral}
\end{equation}
Here and below the application of the $\delta$-function is understood in the sense of the limit in
Eq.~(\ref{Eq:VpsGeneralMel}). The C-tensors $C^{l}\left(  \frac{\nabla}{k_{c}}\right)$ are
differential operators of order $l$, defined as
\begin{equation}
C_{m}^{l}\left(  \frac{\mathbf{\nabla}}{k_{c}}\right)  =\frac{1}{k_{c}^{l}%
}\left[  \left(  l-m\right)  !\left(  l+m\right)  !\right]  ^{1/2}\sum
_{p,q,r}\left(  \frac{1}{\sqrt{2}}\right)  ^{p+q}\frac{1}{p!q!r!}\left(
\nabla_{+1}\right)  ^{p}\left(  \nabla_{-1}\right)  ^{q}\left(  \nabla
_{0}\right)  ^{r} \, .
\label{Eq:C-tensorOfderivatives}
\end{equation}
The spherical components of the nabla operator $\mathbf{\nabla}$ are
expressed in terms of its Cartesian components
\begin{align*}
\nabla_{+1} &  =-\frac{1}{\sqrt{2}}\left(  \frac{\partial}{\partial x}%
+i\frac{\partial}{\partial y}\right) \, , \\
\nabla_{-1} &  =\frac{1}{\sqrt{2}}\left(  \frac{\partial}{\partial x}%
-i\frac{\partial}{\partial y}\right)  \, ,\\
\nabla_{0} &  =\frac{\partial}{\partial z} \, .
\end{align*}
Explicit representation of several differential operators
$C_{m}^{\ell}\left(  \frac{\nabla}{k_{c}
}\right)$ is given in Table~\ref{Tab:Clm}.

\begin{table}[h]
\begin{center}
\begin{tabular}[c]{ll}%
\hline\hline
$\left(  \ell ,m\right)  $ & $k_{c}^{\ell}\,C_{m}^{\ell}\left(  \frac{\nabla}{k_{c}%
}\right)  $\\
\hline
$\left(  0,0\right)  $ & $1$\\
$\left(  1,0\right)  $ & $\partial_{z}$\\
$\left(  1,\pm1\right)  $ & $\frac{1}{\sqrt{2}}\left(  \mp\partial
_{x}+i\partial_{y}\right)  $\\
$\left(  2,0\right)  $ & $\frac{1}{2}\left(  -\Delta+3\partial_{z}^{2}\right)
$\\
$\left(  2,\pm1\right)  $ & $\mp\sqrt{\frac{3}{2}}\left(  \partial_{x}\pm
i\partial_{y}\right)  \partial_{z}$\\
$\left(  2,\pm2\right)  $ & $\frac{1}{2}\sqrt{\frac{3}{2}}\left(  \partial
_{x}\pm i\partial_{y}\right)  ^{2}$ \\ \hline\hline
\end{tabular}
\end{center}
\caption{  Explicit form of operators
$C_{m}^{\ell}\left(  \frac{\nabla}{k_{c}}\right)$ for several values of $\ell$  computed using
Eq.(\ref{Eq:C-tensorOfderivatives}). We use a shorthand
notation for partial derivatives, e.g.,
$\partial_{z}^{2} \equiv \frac{\partial^2}{\partial_{z}^{2}}$.  \label{Tab:Clm} }
\end{table}

We also mention an alternative representation of the C-tensors, which may be
more useful in some applications%
\[
C_{m}^{l}\left(  \frac{\mathbf{\nabla}}{k_{c}}\right)  =\frac{1}{\left(
k_{c}\right)  ^{l}}\sqrt{\frac{\left(  2l-1\right)  !!}{l!}}\left\{
...\left\{  \left\{  \mathbf{\nabla}\otimes\mathbf{\nabla}\right\}
_{1}\otimes\mathbf{\nabla}\right\}  _{2}...\otimes\mathbf{\nabla}\right\}
_{lm}.
\]

By limiting  the summation to pairs of partial waves $s-s$ (resulting $J=0,m=0$),
$s-d$ and $d-s$ (resulting $J=2, m=0$), we
recover the result, Eq.~(\ref{Eq:PSTensorGauge}), for the truncated dipolar pseudopotential.

We would like to emphasize that so far we have not specialized our consideration
to a particular multipole contribution. In fact the original interaction potential
may contain a sum over several multipolar contributions. In the following sections
we specialize our discussion to two limiting cases: isotropic interactions and
weak anisotropic interaction,
where the Born approximation is valid. In the former case we recover literature
results and in the latter (Born) case we show that only $J=L$ component of the compound
tensor remains in the general expansion~(\ref{Eq:VpsGeneral}).

\section{Specialized cases}
\subsection{Isotropic interactions}
\label{Sec:Isotropic}
In this Section we simplify the general pseudo-potential expression, Eq.~(\ref{Eq:VpsGeneral}),  for the case of
isotropic interactions. In this case the various partial waves decouple and the reactance matrix
reads
\[
\left( \mathcal{K}_{lm}^{l^{\prime}m^{\prime}} \right)_\mathrm{iso}=\delta_{ll^{\prime}}
\delta_{mm^{\prime}}\tan\delta_{l}\, ,
\]
 $\delta_{l}$ being  the conventional phase
shifts. This relation can be easily derived by comparing Eq.(\ref{Eq:asmpt})
with the textbook expressions for isotropic scattering.
The expression for the coupling coefficient $T$, Eq.(\ref{Eq:Tcoupling}), simplifies to
\[
T_{J}\left(  ll^{\prime};k_{c}\right)  =-4\pi\frac{\hbar^{2}}{M}%
(2l+1)^{3/2}\left(  -1\right)  ^{l}\frac{\tan\delta_{l}}{k_{c}}\delta
_{J,0}\delta_{ll^{\prime}} ,
\]
where we used
$\sum_{m}\left(  -1\right)  ^{m}C_{l,m;l,-m}^{J0}=\delta
_{J,0}\left(  -1\right)  ^{l}\sqrt{(2l+1)}$. For isotropic interactions
we deal with only $J=0$ (scalar) component of the compound tensor.
This reflects the rotational properties (scalar) of the full isotropic potential.
Further,  we invoke the theorem for addition of
spherical harmonics
\[
\left\{  C^l\left(  \frac{\overleftarrow{\mathbf{\nabla}}}{k_{c}}\right)
\otimes C^{l^{\prime}}\left(  \frac{\overrightarrow{\mathbf{\nabla}%
}}{k_{c}}\right)  \right\}  _{00}=\frac{\left(  -1\right)  ^{l}}%
{\sqrt{(2l+1) }} \, P_{l}\left(  \frac{\overleftarrow{\mathbf{\nabla}}%
}{k_{c}}\cdot\frac{\overrightarrow{\mathbf{\nabla}}}{k_{c}}\right)
\delta_{ll^{\prime}}%
\,
\]
and  arrive at
\begin{equation}
\left(V_{ps}\right)_\mathrm{iso}=2\pi\frac{\hbar^{2}}{\mu}\sum_{l}(2l+1)\left(  -\frac{\tan\delta_{l}%
}{k_{c}}\right) \delta\left(  \mathbf{r}\right)   P_{l}\left(  \frac{\overleftarrow{\mathbf{\nabla}}}{k_{c}%
}\cdot\frac{\overrightarrow{\mathbf{\nabla}}%
}{k_{c}}\right)
\label{Eq:VpsIsotropic} \, .
\end{equation}
Again the application of the $\delta$-function is understood in the sense of taking the limit in
Eq.~(\ref{Eq:VpsGeneralMel}).
This is a result previously derived by Omont~\cite{Omo77} in the context of
interaction of Rydberg electrons with atoms (in that case, the reduced mass $\mu\approx m_{e}$; in our case
$\mu=M/2$). The $l=0$ ($s$-wave) contribution in this sum is the Fermi pseudopotential.
The $l=1$ and $l=2$  ($p$- and $d$-wave) terms were re-derived recently
in Ref.~\cite{IdzCal06}.

\subsection{Anisotropic interactions in the Born approximation}
\label{Sec:Born}
 The derived pseudopotential is based on a microscopic description
 of the collision process. The relevant K-matrix is to be determined from the
 full solution of the Schrodinger equation with the original potential.
If the perturbation imposed by the interaction potential on the free
motion of the particles is sufficiently weak, one may employ the
Born approximation. Of course, the exact K-matrix is not guaranteed to be equal to its Born value. The Born approximation would break, for example, near resonances, found
in dipolar scattering in Ref.~\cite{DebYou01}. Even without resonances, according to Ref.~\cite{Wan07}, the
Born approximation for dipolar scattering is valid only if $\bar{R}_3/R_\mathrm{vdW} \ll 1$, where
$R_\mathrm{vdW} \sim 100 \, \mathrm{bohr}$ is the van der Waals
length ($\bar{R_6}$ based
on the van der Waals dispersion coefficient $C_6$). This inequality easily breaks for polar molecules, where $\bar{R}_3 \sim 10^4 \, \mathrm{bohr}$.

Nevertheless, it is instructive to consider the pseudopotential in the Born regime.
To the lowest order, the
K-matrix in the Born approximation for anisotropic interactions was derived
in Ref.~\cite{Der03}. It reads
\begin{align}
\left(  \mathcal{K}_{lm}^{l^{\prime}m^{\prime}}\right)  _{\mathrm{Born}} &
=-\frac{M}{\hbar^{2}}I_{ll^{\prime}}^{\left(  L\right)  }~\langle
lm\left\vert C_{0}^{L}\right\vert l^{\prime}m^{\prime}\rangle,\\
I_{ll^{\prime}}^{\left(  L\right)  } &  =\int_{0}^{\infty} x^2 dx V_L(x)
j_{l}\left( k x\right)  j_{l^{\prime}}\left( k  x\right)  .
\end{align}
Here $\langle lm\left\vert C_{0}^{L}\right\vert l^{\prime}m^{\prime
}\rangle=\int d\Omega Y_{lm}^{\ast}\left(  \Omega\right)  C_{0}^{L}\left(
\Omega\right)  Y_{l^{\prime}m^{\prime}}\left(  \Omega\right)  $.  Invoking the
Wigner-Eckart theorem for this matrix element of the C-tensor, we may
introduce the reduced matrix element of the K-matrix, $\langle l||\mathcal{K}^{\left(
L\right)  }||l^{\prime}\rangle$, via
\begin{equation}
\mathcal{K}_{lm}^{l^{\prime}m^{\prime}}=(-1)^{l-m}\left(
\begin{array}
[c]{ccc}%
l & L & l^{\prime}\\
-m & 0 & m^{\prime}%
\end{array}
\right)  \langle l||\mathcal{K}^{\left(  L\right)  }||l^{\prime}\rangle \, ,
\label{Eq:KmatBornWET}
\end{equation}
\begin{equation}
\langle l||\mathcal{K}^{\left(  L\right)  }||l^{\prime}\rangle=-\frac{M}{\hbar
^{2}}I_{l^{\prime}l}^{\left(  L\right)  }\langle l||C^{\left(  L\right)
}||l^{\prime}\rangle \, ,
\end{equation}
with the reduced matrix element of the C-tensor being conventionally expressed
as%
\begin{equation}
\langle l||C^{\left(  L\right)  }||l^{\prime}\rangle=(-1)^{l}\;\sqrt{\left(
2l+1\right)  \left(  2l^{\prime}+1\right)  }\;\left(
\begin{array}
[c]{ccc}%
l & L & l^{\prime}\\
0 & 0 & 0
\end{array}
\right)  .
\end{equation}
By substituting Eq.(\ref{Eq:KmatBornWET}) into Eq.(\ref{Eq:Tcoupling}),
summing over magnetic quantum numbers, and  using the orthogonality relation
for the Clebsh-Gordan coefficients, we obtain the simplified expression for
the coupling coefficient
\begin{equation}
T_{J}\left(  ll^{\prime};k_{c}\right)  =-4\pi\frac{\hbar^{2}}{M}\left(
\frac{(2l+1)(2l^{\prime}+1)}{2L+1}\right)  ^{1/2}\left(  -1\right)
^{l^{\prime}}\frac{\langle l||K^{\left(  L\right)  }||l^{\prime}\rangle}%
{k_{c}}\delta_{L,J}.
\end{equation}
Accordingly, only the $J=L$ compound tensor participates in the pseudopotential%
\begin{align}
\left(  V_{ps}\right)  _{\mathrm{Born,}L}  & =-4\pi\frac{\hbar^{2}}{M}%
\sum_{ll^{\prime}}(-1)^{l^{\prime}}\left(  \frac{(2l+1)(2l^{\prime}+1)}%
{2L+1}\right)  ^{1/2}\frac{\langle l||\mathcal{K}^{\left(  L\right)
}||l^{\prime}\rangle}{k_{c}}\times  \nonumber \\
& \delta\left(  \mathbf{r}\right) \left\{  C^{l}\left(  \frac{\overleftarrow{\mathbf{\nabla}}}{k_{c}}\right)
\otimes  C^{l^{\prime}}\left(  \frac
{\overrightarrow{\mathbf{\nabla}}}{k_{c}}\right)  \right\}  _{L0} \, .
\label{Eq:PseudoBornL}
\end{align}

For the specific case of dipolar interactions, $L=2$,  selection rules require $l'=l$ or $l'=l\pm2$.
The reduced matrix elements of the K-matrix are proportional to the wave-vector (c.f. introduction
of generalized scattering lengths for dipolar scattering in Ref.~\cite{Der03}). Explicitly,
\begin{equation}
\langle l||K^{\left(  L=2\right)  }||l^{\prime}\rangle_{\mathrm{Born,DD}%
}=-D^{2}\frac{M}{\hbar^{2}}k~\langle l ||C^{2}|| l^{\prime
}\rangle\times\left\{
\begin{array}
[c]{cc}%
\left[  l\left(  l+1\right)  \right]  ^{-1} & l^{\prime}=l\\
\left[  3\left(  l+1\right)  \left(  l+2\right)  \right]  ^{-1} & l^{\prime
}=l+2\\
\left[  3l\left(  l-1\right)  \right]  ^{-1} & l^{\prime}=l-2
\end{array}
\right. \,.
\end{equation}
By limiting the summation to  $l=0,l'=2$ and $l=2,l'=0$ couplings in Eq.(\ref{Eq:PseudoBornL}),
 we recover the truncated
dipolar pseudopotential, Eq.(\ref{Eq:PSTensorGauge}), where we use
the off-diagonal scattering length in the Born approximation~\cite{Der03}
\begin{equation}
 \left(a_{sd}\right)_\mathrm{Born} = -\frac{1}{6\sqrt{5}}D^{2}\frac{M}{\hbar^{2}} \, .
\end{equation}
Notice that $\left(a_{sd}\right)_\mathrm{Born}$ is of the same order of magnitude
as the radius of the asymptotic region for the pseudopotential, $\bar{R}_3$, Eq.~(\ref{Eq:Rmulti}).

Due to the restriction to the $J=L, M=0$ component of the compound tensor,
the transformational properties of the pseudopotential under rotations of the coordinate frame
are the same as of the original $V_L$-component of the potential. While we have proven this property
in the Born approximation, it seems not to hold in the non-perturbative regime.
For example, in the case of dipolar interactions,
radial equations for the $m=0$ component of the scattering wavefunction involve
couplings of $s-d$, $d-g$, $g-h$, ... partial waves, while for the $m= \pm 2$ components,
the $s-d$ coupling is missing. A resonance in the s-wave channel would drastically
affect the K-matrix for the $m=0$ channels, while the $m=  \pm 2$ components are
not affected by this resonance.  In this case, the Wigner-Eckart theorem for
the K-matrix, Eq.~(\ref{Eq:KmatBornWET}), does not hold and we no longer arrive
at the restriction $J=L$ in the pseudopotential.

Finally, it is worth pointing out the connection of the derived Born formulae
to the original potential, Eq.(\ref{Eq:VL-general}). This also provides 
another test of self-consistency of the derivation. To this end, we transform
the original potential to the momentum-representation
\[
  v(\mathbf{k'},\mathbf{k}) = \frac{1}{(2\pi)^3} \, \int d\mathbf{r} e^{-i\mathbf{k}\cdot \mathbf{r}}
  V_L(r)\, P_L(\hat{\mathbf{r}})  e^{i\mathbf{k}\cdot \mathbf{r}} \,.
\]
By expanding the plane waves into the spherical waves, evaluating angular integrals,
and setting the $|\mathbf{k}|=|\mathbf{k'}|=k_c$ (on-shell values)
we arrive at the pseudopotential kernel, Eq.(\ref{Eq:kernel}), with the elements of the
K-matrix evaluated in the Born approximation. This means that following
the derivation of this paper we would necessarily arrive at the pseudopotential
in the Born-approximation,
Eq.(\ref{Eq:PseudoBornL}), even discarding the microscopic description of the collision physics.
This is another test of self-consistency of the present approach in the Born approximation.
We emphasize again that  a description starting from Eq.~(\ref{Eq:VL-general}) would miss deviations of the scattering matrix from it's Born value.

\section{Summary}
\label{Sec:Summary}
To reiterate, here we derived a contact pseudo-potential for
anisotropically interacting particles trapped in a harmonic potential.
Such interactions include, for example, dipole-dipole interactions between polarized
molecules and atoms and subsume the conventional isotropic interactions.
Due to the well-defined  collision energy and assumption of sufficiently
smooth wave-functions, the previously derived~\cite{Der03}
rigorous, but complicated, pseudopotential is simplified.
An additional assumption is that the characteristic radius of
the potential, $\bar{R}_k$ is much smaller than the harmonic length of the trapping
potential. All of these assumptions are natural in the contexts of studies of
many-body physics, e.g., BECs of dipolar gases.

The simplified contact pseudopotential, Eq.~(\ref{Eq:VpsGeneral}), involves
a sum over partial derivatives. The anisotropy of the
interaction is reflected in the preferential appearance
of derivatives along the polarizing field. Compared to the rigorous
pseudo-potential of Ref.~\cite{Der03},
the presented simplified
pseudo-potential may prove to be more practical in applications to
trapped quantum gases of polarized atoms and molecules. The advantage
of the pseudo-potential description is that, being based on the
microscopic  quantum description of the collision process, it remains valid even in
the non-perturbative regime, where the Born approximation breaks down.

I would like to thank E. Tiesinga and P. Naidon for discussions and C. Greene
for bringing Ref.~\cite{Omo77} to my attention. This
work was supported in part by the  NSF and NASA-EPSCOR program.

%\bibliography{all}

\end{document}